\documentstyle[epsfig]{aipproc}
\def\etal{{\it et al.}}
 
\begin{document}
\title{The High Time Resolution Spectral Evolution of Gamma-Ray Bursts} 

\author{David L. Band$^*$ and Lyle A. Ford$^{\dagger}$} 
\address{$^*$CASS, UC San Diego, La Jolla, CA  92093 \\
$^{\dagger}$University of Wisconsin, Eau Claire, WI 54702}

\maketitle

\begin{abstract}
Previous studies of the evolution of gamma-ray burst spectra have
generally relied on fitting sequences of spectra.  These studies
usually were confined to bright bursts and often lacked sufficient
temporal resolution.  We have developed techniques which sacrifice
spectral resolution for temporal resolution.  First, the
crosscorrelations between intensity lightcurves in different energy
bands allow for the classification of the spectral evolution of a
large burst sample. Second, the energy correlation of pairs of counts
places limits on short-duration, narrowband emission. 
\end{abstract}
\section*{Introduction}
Since the observed gamma-ray burst spectrum reflects the physical
processes within the source's emitting region, spectral variations are
an important diagnostic of the nature of this region. Studies showed
that the spectrum hardens during intensity spikes, but there is a
hard-to-soft trend during these spikes, and the hardness tends to peak
at successively lower values from spike to spike (e.g., using
SIGNE\cite{kargatis94} and BATSE \cite{ford95} spectra). These studies
fitted sequences of spectra and then compared the lightcurves of the
intensity and a measure of the spectral hardness derived from these
fits (e.g., the energy $E_p$ of the peak of $E^2 N(E)\propto \nu
F_\nu$). Since the spectrum must have sufficient counts for a reliable
fit, these studies often lacked the temporal resolution to determine the
spectral evolution of the structure which is evident in the intensity
lightcurves, and the best temporal resolution was available only for
the brightest bursts. 

The BATSE spectra accumulated over timescales as short as 0.1~s are
well described by the broadband ``GRB'' spectral model consisting of
two power laws which join smoothly\cite{band93}.  Yet, early models of
a cosmological fireball predicted a narrowband, quasi-black body
spectrum when the fireball becomes optically thin
\cite{goodman86,paczynski86}.  Fireball models have developed greatly
in the past decade, and they no longer predict narrowband spectra.
Nonetheless, the question remains whether burst spectra are narrowband
on short timescales. 

BATSE provides burst data with a variety of temporal and spectral
resolutions; usually there is a tradeoff between the two.  Here we
emphasize temporal resolution at the expense
of spectral resolution.  First, we used the auto- and crosscorrelation
of lightcurves in different energy bands to categorize the evolution
in a large burst sample \cite{band97}. Second, we searched for short
duration, narrowband emission by correlating the energies of pairs of
counts as a function of the arrival time separations\cite{ford96}.  
This methodology uses individual counts identified by arrival time and
energy; the number of such counts is insufficient for forming a
spectrum on millisecond timescales. 
\section*{Auto- and Crosscorrelations of Energy Channels} 
To characterize the spectral evolution of a large sample of bursts we
use the auto- and crosscorrelation functions (ACF and CCFs,
respectively) of burst lightcurves in different energy channels
\cite{band97}. The BATSE LADs provide discriminator rates in 4 energy 
bands (Ch.~1: 25--50, Ch.~2: 50--100, Ch.~3: 100--300, and Ch.~4:
300--2000~keV) on a 64~ms timescale before, during and after a burst.
Most temporal analysis assumes that the time series being analyzed is
a sample from a stationary process, which justifies the use of
periodic functions in the analysis. However, bursts are transient
phenomena; ACFs and CCFs do not assume that the signal is stationary. 

The ACFs and CCFs of transient events such as bursts involve some
subtleties \cite{fenimore95,band97}. Instead of using a 
zero-mean time series (i.e., the mean value is subtracted from the
lightcurve), we use a background-subtracted time series, which is
equivalent to a zero-mean time series of a burst with infinite
stretches of background before and after the burst.  Assuming the
noise is dominated by Poisson statistics, we can correct for this
noise by modifying the variance. The resulting ACFs and CCFs are very
robust, and give the correct shape even for signal-to-noise ratios of
order unity at the peaks of the lightcurve.  Only when there is little
or no apparent signal, as is sometimes the case for the highest energy
channel ($E>300$~keV), is the CCF so noisy as to be unusable. 

\begin{figure}[b!] 
\centerline{\epsfig{file=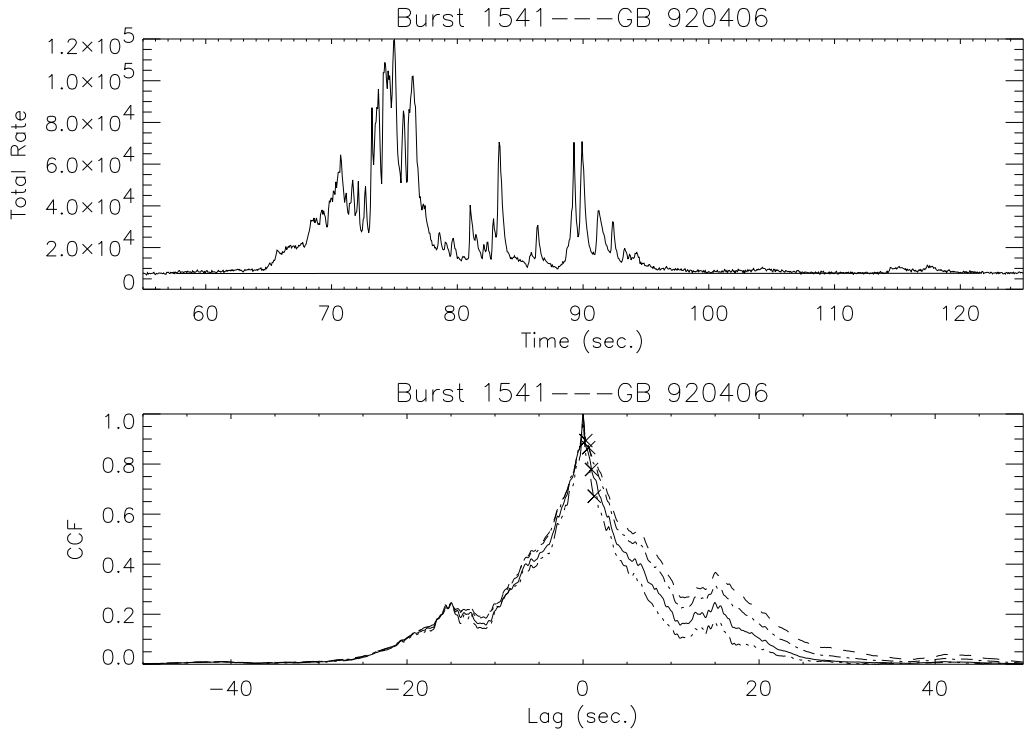,width=5.5in}}
\caption{Lightcurve (top) and the ACF and CCFs (bottom) of GRB~920406.
The solid curve is the ACF of the fiducial energy
channel (Ch.~3), the 3~dots-dashed curve is the CCF with a higher
energy channel (Ch.~4), and the two remaining curves are the CCFs with
lower energy channels (Ch.~1---dashed---and Ch.~2---dot-dashed).  The
order of the correlation curves indicate the type of spectral
evolution.  Specifically, for hard-to-soft evolution the order (from
top to bottom) of the curves on the positive lag side is:  dashed,
dot-dashed, solid and 3 dots-dashed. }
\label{fig1}
\end{figure}
We calculate the CCFs of a fiducial energy channel, Ch.~3
(100--300~keV), with each of the 4 energy channels (the CCF of the
fiducial channel with itself is that channel's ACF).  Through
simulations and analytic modeling we have developed diagnostics for
different types of spectral evolution.  We compare the time lags of
the peaks of each curve and their relative values at different lags,
as shown by the example in Figure~1. 

We calculated the ACFs and CCFs for 209 strong, mostly long bursts
\cite{band97}.  The order of the CCF peaks shows that in general high
energy emission precedes low energy emission. As was
known previously from comparing the ACFs of the different channels
\cite{fenimore95}, the CCF widths indicate that high energy
temporal structure is narrower than low energy structure (i.e., spikes
last longer at low energy than at high).  The relative
order of the CCFs at different lags shows there is
hard-to-soft evolution within and among spikes in $\sim$80--90\% of
the bursts, and there are only a few cases of soft-to-hard spectral
evolution.  The peaks of the CCFs for the high
energy channels typically lead those of the low energy channels by
0.1-0.2~s, and the CCFs of the low energy channels are typically 
$\sim$25\% broader (where the CCFs have a value of 0.5) than those of
the high energy channels.  Thus this study shows that
hard-to-soft spectral evolution is ubiquitous but counterexamples
exist. 
\section*{Broadband vs. Narrowband Emission} 
In burst spectra the photons are distributed over a broader energy
band than a simple black body spectrum.  The question is whether the
spectrum is inherently broadband or is composed of short duration
narrowband (e.g., black body) events which rapidly sum to a broadband
spectrum.  Narrowband spectra may indicate thermal processes. BATSE
provides full spectra only on timescales of 0.128~s or longer; spectra
can be formed on shorter timescales from the SD time-tagged events
(the STTE datatype provides a list of 64,000 counts with their arrival
time to 128$\mu$s and energy in one of 256 channels), but the photon
flux is insufficient to accumulate spectra on timescales of 1--10~ms.
Consequently we searched for short duration
narrowband events by correlating the energies of pairs of
counts\cite{ford96}.  If there is short duration narrowband emission,
then the energies of count pairs with a small arrival time separation
should be more correlated than counts with a large separation. 

Thus we would like the fraction of count pairs whose energies differ
by less than a certain fraction of their average energy (i.e., those
pairs for which $|E_1-E_2|<f(E_1+E_2)/2$).  However, since the
detectors are imperfect, the counts assigned to a given apparent
energy bin could have originated from a broad energy range.  With
assumptions about the incoming photon spectrum, the detector response
can be inverted to provide a distribution $p(E \,|\, E^\prime)$ of the
likely photon energy $E$ given the apparent energy $E^\prime$ (the
energy reported by the detector).  These distributions are used to
calculate the likely probability the two counts had energies within a
certain fractional separation: 
\begin{equation}
L(E_1^\prime,E_2^\prime) = \int_0^\infty dE 
   \int_{-fE/(1+f)}^{fE/(1-f)} d(\Delta E) \,
   p(E\,|\,E_1^\prime) p(E+\Delta E\,|\,E_2^\prime) \quad .
\end{equation}
This probability $L(E_1^\prime,E_2^\prime)$ is then averaged over all
pairs of counts with a separation $\Delta t$, resulting in a function
$A(\Delta t)$. 

\begin{figure}[b!] 
\centerline{\epsfig{file=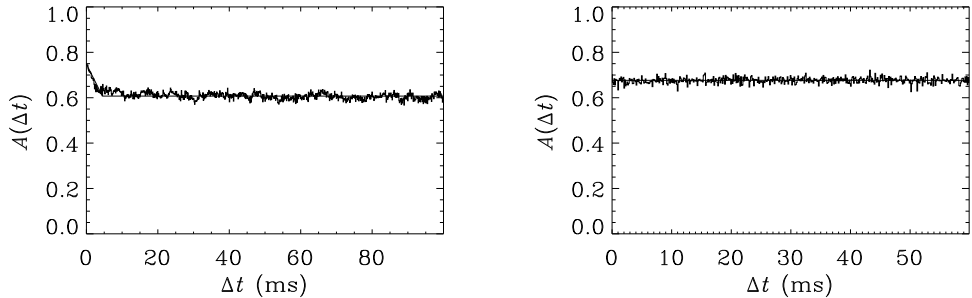,width=5.5in}}
\caption{The energy separation statistic $A(\Delta\tau)$ as a function
of the time separation.  On the left is a simulation where the spectrum
is formed from black bodies with a distribution of temperatures and
durations, while on the right is data from GRB~940717.  Note the upturn
in $A(\Delta\tau)$ for the simulation with narrowband emission, and the 
absence of an upturn for the actual data.} 
\label{fig2}
\end{figure}
This function $A(\Delta t)$ should be constant if the spectrum is
constant, but $A(\Delta t)$ will increase for decreasing $\Delta t$ if
there is short duration narrowband emission, as has been verified by
simulations (see the lefthand side of Figure~2).  This methodology
will identify narrowband emission superimposed on a broadband
spectrum. We are quantifying the fraction of narrowband emission on a
given timescale which is consistent with the data; we anticipate that
this fraction will usually be an upper limit. 

This methodology has been applied to 20 bright, short duration (less
than 2~s) bursts.  No evidence of short duration narrowband emission
was found (see the right-hand side of Figure~2 for an example), with
upper limits on the fraction of such emission of order 10\% on
timescales down to 1~ms. A
complication which needs to be solved is the normal spectral evolution
on timescales longer than we are interested. 

B.~Schaefer (private communication, 1997) is undertaking a similar
study using the LAD's Time Tagged Events (TTE) datatype which provides
16,000 counts in 4 energy channels to 2$\mu$s accuracy. 
\section*{Acknowledgments} 
The work of the UCSD group is supported by NASA contract NAS8-36081.

\end{document}